\documentstyle[12pt]{article}
\begin{document}
\thispagestyle{empty}
\begin{flushright}
JINR E2-93-159
\end{flushright}
\begin{center}
\Large{A CLOSED EXPRESSION\\FOR THE UNIVERSAL $\cal R$-MATRIX\\
IN A NON-STANDARD QUANTUM DOUBLE}
\end{center}

\vspace{.5cm}

\begin{center}
\large{A.A.VLADIMIROV}${}^{\,*\,\diamond}$
\end{center}
\begin{center}
\large{Laboratory of Theoretical Physics, \\
Joint Institute for Nuclear Research, \\
Dubna, Moscow region, 141980, Russia}
\end{center}

\vspace{1cm}

\begin{center}
ABSTRACT
\end{center}

In recent papers of the author, a method was developed for constructing
quasitriangular Hopf algebras (quantum groups) of the quantum-double
type. As a by-product, a novel non-standard example of the quantum
double has been found. In the present paper, a closed expression (in
terms of elementary functions) for the corresponding universal
$\cal R$-matrix is obtained. In reduced form, when the number of
generators becomes two instead of four, this quantum group can be
interpreted as a deformation of the Lie algebra $[x,h]=2h$ in the
context of Drinfeld's quantization program.

\vspace{4cm}

${}^*$ Work supported in part by the Russian Foundation of Fundamental
Research (grant 93-02-3827)

\vspace{.3cm}

${}^\diamond$ E-mail: alvladim@theor.jinrc.dubna.su

\pagebreak

In papers~\cite{Vl1,Vl2} we have modified the recipes of~\cite{FRT,Ma}
and developed a regular method for constructing a quantum double
out of any invertible constant matrix solution
$R$ of the quantum Yang-Baxter equation (QYBE)
\begin{equation}
R_{12}R_{13}R_{23}=R_{23}R_{13}R_{12}\,.  \label{1}
\end{equation}
To illustrate the efficiency of the method, an $R$-matrix
from the two-parameter class
\begin{equation}
\left( \begin{array}{cccc}1&p&-p&pq\\0&1&0&q\\0&0&1&-q\\0&0&0&1
\end{array}      \right)  \label{2}
\end{equation}
discovered by D.Gurevich (cited in~\cite{Ly}) and studied also in
[6--12], has been taken as an input (actually, with $p=q=1$). The
result~\cite{Vl2} is a new non-standard quantum double with four
generators $\{b,g,v,h\}$ obeying the following relations:
$$ [g,b]=[h,b]=2\sinh g\,,\ \ \ [g,v]=[h,v]=-2\sinh h\,, $$
$$ [b,v]=2(\cosh g)v+2(\cosh h)b\,,\ \ \ \ [g,h]=0\,, $$
\begin{equation}
\Delta(b)=e^g\otimes b+b\otimes e^{-g}\,,\ \ \
\Delta(v)=e^h\otimes v+v\otimes e^{-h}\,, \label{3}
\end{equation}
$$ \Delta(g)=g\otimes 1+1\otimes g\,,\
\Delta(h)=h\otimes 1+1\otimes h\,,\ S^{\pm1}(g)=-g\,, $$
$$ S^{\pm1}(h)=-h\,,\ S^{\pm1}(b)=-b\pm2\sinh g\,,\
S^{\pm1}(v)=-v\mp2\sinh h\,. $$

A month later, Burdik and Hellinger~\cite{BH} introduced a quantum
double also related to $R$-matrix (\ref{2}) in terms of generators
$\{\tau ,\pi ,T,P\}$ and a parameter $\gamma $. It is not difficult to
verify that their double is isomorphic to (\ref{3}) due to the
following identification:
\begin{equation}
\tau =e^g\,b\,,\ \ \pi =\frac{1-e^{-2g}}{\gamma}\,,\ \
T=h\,,\ \ P=\frac{\gamma }{2}\,e^h\,v\,. \label{4}
\end{equation}

The universal $\cal R$-matrix of the quantum double (\ref{3})
is displayed in~\cite{Vl2} as several terms of its power expansion in
$g$ and $h$ (in~\cite{BH} -- as a power series in appropriately chosen
combinations of generators). The main result of the present paper is an
explicit formula for $\cal R$:
\begin{equation} {\cal
R}=\exp\left\{\frac{g\otimes 1+1\otimes h} {\sinh(g\otimes 1+1\otimes
h)}(\sinh g\otimes v+b\otimes \sinh h) \right\}\,. \label{5}
\end{equation}
This has been guessed with the use of computer (namely, the
symbolic calculation program FORM~\cite{Ve}) and then proved by hand.
I believe that expanding (\ref{5}) and taking (\ref{4}) into account
should eventually yield the power-series expression for $\cal R$ given
in \cite{BH}.

The key property of $\cal R$ to be proved is its
quasicocommutativity~\cite{Dr2}. For example, the $\cal R$-matrix
(\ref{5}) must obey
\begin{equation} {\cal R}(e^h\otimes v+v\otimes
e^{-h}){\cal R}^{-1}= e^{-h}\otimes v+v\otimes e^h\,.  \label{6}
\end{equation}
Denoting ${\cal R}=\exp A$ we come to
\begin{equation}
2(v\otimes \sinh h-\sinh h\otimes v)=[A,\Delta(v)]+
\frac{1}{2}[A,[A,\Delta(v)]]+\ldots \,,  \label{7}
\end{equation}
as it follows from the Hadamard formula. Denoting also
\begin{equation}
\Phi=\frac{z}{\sinh z}\,,\ \ \Phi'=\frac{d}{dz}\,\frac{z}{\sinh z}\ \
\ \ {\rm with} \ \ \ z=g\otimes 1+1\otimes h\,,  \label{8}
\end{equation}
we find
\begin{equation}
[A,\Delta(v)]=2(v\otimes \sinh h-\sinh h\otimes v)-
2(\Phi+\Phi')D\,, \label{9}
\end{equation}
$$D=\sinh(g\otimes 1+1\otimes h)\,(v\otimes \sinh h-\sinh h\otimes v)$$
\begin{equation}
+\sinh(h\otimes 1+1\otimes h)\,(\sinh g\otimes v+b\otimes \sinh h)\,.
\label{10}  \end{equation}
 From the relations
\begin{equation}
[g\otimes 1+1\otimes h\,,\sinh g\otimes v+b\otimes \sinh h]=0\,,
\label{11} \end{equation}
\begin{equation}
[g\otimes 1+1\otimes h\,,\sinh h\otimes v-v\otimes \sinh h]=0\,,
\label{12}  \end{equation}
\begin{equation}
[g\otimes 1+1\otimes h\,,D]=0\,,  \label{13}
\end{equation}
\begin{equation}
[\sinh g\otimes v+b\otimes \sinh h\,,D]=2\sinh(g\otimes 1+1\otimes h)D
\,,  \label{14}  \end{equation}
we deduce
\begin{equation}
[A,\Phi]=[A,\Phi']=[D,\Phi]=[D,\Phi']=0\,,  \label{15}
\end{equation}
\begin{equation}
[A\,,v\otimes \sinh h-\sinh h\otimes v]=2\Phi D\,,  \label{16}
\end{equation}
\begin{equation}
[A,D]=2(g\otimes 1+1\otimes h)D\,.  \label{17}
\end{equation}
The last equality enables us to keep multiple commutators in (\ref{7})
under control and sum them up, with a desired result.

There is no need of a special proof of the other requirements on $\cal
R$~\cite{Dr2}, because an iterative solution of (\ref{6}) is unique in
the Hopf algebra (\ref{3}). Therefore, the universal $\cal R$-matrix
(\ref{5}) obeys QYBE.

\vspace{.5cm}

It is also interesting to consider the reduced version of (\ref{3}),
that is the Hopf algebra with generators $\{v,h\}$ and relations
$$ [v,h]=2\sinh h\,, $$
\begin{equation}
\Delta(v)=e^h\otimes v+v\otimes e^{-h}\,,\ \ \
\Delta(h)=h\otimes 1+1\otimes h\,,  \label{18}
\end{equation}
$$ S^{\pm1}(h)=-h\,,\ \ \ S^{\pm1}(v)=-v\mp2\sinh h\,. $$
Algebra (\ref{18}) is a subalgebra of (\ref{3}) and, at the same time,
the quotient algebra with respect to the centre of (\ref{3}).
The latter is generated by the elements
\begin{equation}
\{\ \ h\!-\!g\,,\ \ (\sinh g)v+(\sinh h)b\ \ \}\,.  \label{21}
\end{equation}
Simply speaking, (\ref{3}) reduces to (\ref{18}) by means of a
substitution
\begin{equation}
g=h\,,\ \ \ \ \ b=-v\,.  \label{19}
\end{equation}

Another way to get (\ref{18}) is to begin with the $R$-matrix (\ref{2})
and use the original Majid's procedure~\cite{Ma}, instead of the above
one~\cite{Vl1,Vl2}, to build a quasitriangular Hopf algebra. Recall
\cite{Vl1} that Majid's approach is based on the $<T,L^{\pm}>=R^{\pm}$
duality whereas we proceed from $<L^-,L^+>=R^{-1}$. In the $sl_q(2)$
case both procedures lead to the same result~\cite{Vl1,Ma}, but in the
case (\ref{2}), due to $R^+\equiv R_{12}=R_{21}^{-1}\equiv R^-$, the
resulting Hopf algebras are substantially different.

By construction, the Hopf algebra (\ref{18}) is quasitriangular (but is
not a quantum double, of course). Its universal $\cal R$-matrix is
obtained by substituting (\ref{19}) into (\ref{5}) and looks like
\begin{equation}
{\cal R}=\exp\left\{\Delta\!\left(\frac{h}{\sinh h}\right)
(\sinh h\otimes v-v\otimes \sinh h)\right\}\,. \label{20}
\end{equation}
By the way, to prove (\ref{20}) directly is easier than
(\ref{5}) because $[A,[A,\Delta(v)]]$ in eq. (\ref{7}) vanishes in
this case.

It is worth mentioning that the standard matrix format for an algebra
(\ref{18}) admits, analogously to $sl_q(2)$ [15-17], an exact
exponential parametrization:
\begin{equation}
\left( \begin{array}{cc}e^h&v\\0&e^{-h} \end{array}\right)=
\exp\left( \begin{array}{cc}h&y\\0&-h \end{array}\right)\,,\ \ \
[y,h]=2h\,,  \label{22}
\end{equation}
where
\begin{equation}
v=\frac{\sinh h}{h}y+\cosh h-\sinh h-\frac{\sinh h}{h}\,. \label{23}
\end{equation}

A similar reparametrization,
\begin{equation}
v=\frac{\sinh h}{h}x\,, \label{30}
\end{equation}
transforms (\ref{18}) into a Hopf algebra
\begin{equation}
[x,h]=2h\,,  \label{24}
\end{equation}
\begin{equation}
\Delta(h)=h\otimes 1+1\otimes h\,,   \label{25}
\end{equation}
\begin{equation}
\Delta(x)=\Delta\!\left(\frac{h}{\sinh h}\right)\left(e^h\otimes
\frac{\sinh h}{h}x+\frac{\sinh h}{h}x\otimes e^{-h}\right)\,,
\label{26}                \end{equation}
\begin{equation}
S^{\pm1}(h)=-h\,,\ \ \ S^{\pm1}(x)=-x+2\left(h\frac{e^{\mp h}}{\sinh h}
-1\right)\,,  \label{27}
\end{equation}
which can be viewed as a deformation of the universal enveloping
algebra of (\ref{24}) treated as a (trivial) Hopf algebra
\begin{equation}
[x,h]=2h\,,\ \ \Delta_0(h)=h\otimes 1+1\otimes h\,,\ \
\Delta_0(x)=x\otimes 1+1\otimes x\,,  \label{28}
\end{equation}
\begin{equation}
S_0(h)=-h\,,\ \ \ \ S_0(x)=-x\,.  \label{29}
\end{equation}
The universal $\cal R$-matrix takes the form
$$ {\cal R}=\exp\left\{\Delta\!\left(\frac{h}{\sinh h}\right)
\left(\frac{\sinh h}{h}\otimes \frac{\sinh h}{h}\right)
(h\otimes x-x\otimes h)\right\} $$
\begin{equation}
=1\otimes 1+h\otimes x-x\otimes h+{\cal O}(h^2)\,.  \label{31}
\end{equation}
According to Drinfeld~\cite{Dr3}, this can be interpreted as the
quantization (with $\hbar=1$) of the classical $r$-matrix
\begin{equation}
r=h\otimes x-x\otimes h\,.  \label{32}
\end{equation}
It is proved in~\cite{Dr3} that such a quantization exists and is
unique. Our relations (\ref{26}), (\ref{27}) and (\ref{31}) produce
it in an explicit form.

Universal $\cal R$-matrix (\ref{31}) obeys QYBE (\ref{1}) in an
abstract algebra (\ref{24}) as well as in all its representations.
For instance, to recover the $R$-matrix (\ref{2}) with $p=q=1$, one has
to substitute into (\ref{31}) the $2\times2$-matrices
\begin{equation}
x=\left(\begin{array}{cc}1&0\\0&-1 \end{array}\right)\,,\ \ \ \ \
h=\left(\begin{array}{cc}0&-1\\0&0 \end{array}\right)\,.  \label{33}
\end{equation}

In conclusion we should remark that in~\cite{Oh}, where the problem
of quantizing (\ref{24}) was also studied, an explicit formula has been
written for an invertible element $F$ which, according
to~\cite{Dr3}, deforms the coproduct,
\begin{equation}
\Delta(x)=F\Delta_0(x)F^{-1}\,,  \label{34}
\end{equation}
and is related to universal $\cal R$-matrix by
\begin{equation}
{\cal R}_{12}=F_{21}F_{12}^{-1}\,.  \label{35}
\end{equation}
However, a straightforward calculation shows that the r.h.s. of
(\ref{35}) with $F$ given in~\cite{Oh} neither coincides with (\ref{31})
nor obeys QYBE (\ref{1}).

An open question is whether $\cal R$ (\ref{31}) (and maybe also
$F$ in closed form) can be obtained by the very interesting direct
method recently proposed~\cite{FM} for evaluating quantum objects like
$\cal R$ and $F$ as functionals of the corresponding classical
$r$-matrix.

\vspace{.5cm}

\noindent {\bf Acknowledgments}

\vspace{.3cm}

\noindent
I wish to thank L.Avdeev, I.Aref'eva, L.Faddeev, A.Isaev, P.Kulish,
V.Lyakhovsky, V.Priezzhev, P.Pyatov and V.Tolstoy for helpful
discussions.

\end{document}